\documentclass[reprint,amsmath,amssymb,
 aps,prl,twocolum,superscriptaddress]{revtex4-1}
\usepackage{graphicx}
\usepackage{dcolumn}
\usepackage{bm}

\begin{document}


\title{Hydrodynamic fluctuations in confined emulsions}



\author{Nicolas Desreumaux}
\email[]{nicolas.desreumaux@espci.fr}
\author{Jean-Baptiste Caussin}
\email[]{jeanbaptiste.caussin@ens-lyon.fr}
\author{Raphael Jeanneret}
\email[]{raphael.jeanneret@espci.fr}
\affiliation{Laboratoire de physique et m\'ecanique des milieux h\'et\'erog\`enes, 
PMMH, ESPCI ParisTech, CNRS UMR 7636, Universit\'e Paris 6 et Universit\'e Paris 7, 10 rue Vauquelin, 75005 Paris, France.}
\author{Eric Lauga}
\email[]{elauga@ucsd.edu}
\affiliation{Department of Mechanical and Aerospace Engineering, University of California San Diego, 9500 Gilman Drive, La Jolla CA 92093-0411, USA.}
\author{Denis Bartolo}
\email[]{denis.bartolo@ens-lyon.fr}
\affiliation{Laboratoire de physique et m\'ecanique des milieux h\'et\'erog\`enes, 
PMMH, ESPCI ParisTech, CNRS UMR 7636, Universit\'e Paris 6 et Universit\'e Paris 7, 10 rue Vauquelin, 75005 Paris, France.}
\affiliation{Laboratoire de Physique de l'Ecole Normale Sup\'erieure de Lyon, Universit\'e de Lyon and CNRS, 46, all\'ee d'Italie, F-69007 Lyon, France.}


\date{\today}

\begin{abstract}
When an ensemble of particles interact hydrodynamically, they generically display large-scale transient structures such as swirls in sedimenting particles~\cite{chaikin},  or colloidal strings in sheared suspensions~\cite{itai}. Understanding  these nonequilibrium fluctuations is a very difficult problem, yet they are  of great importance for a wide range of processes including pigment deposition in cosmetic or paint  films, the transport of microfluidic droplets, ... All these samples concern rigidly confined fluids, which we consider in this paper.  
We address the collective dynamics of non-Brownian droplets cruising in a shallow microchannel. We provide a comprehensive characterization of their  spatiotemporal density fluctuations. We  show that density excitations freely propagate at all scales, and in all directions even though the particles are neither affected by potential forces nor by inertia.
We introduce a theory which quantitatively accounts for our experimental findings. By doing so we demonstrate that the fluctuation spectrum of this nonequilibrium system is shaped by the combination of truly long-range hydrodynamic interactions and local collisions.

\end{abstract}


\maketitle

Understanding the collective dynamics of non-brownian particles in  viscous fluids is a long-standing challenge in fluid mechanics.   A typical example is the still unsolved problem of  sedimentation in a quiescent fluid.  Rather than falling along straight lines, as an isolated particle would do, sedimenting particles experience additional swirling motion correlated over  large yet finite distances. The physical origin of those dynamical structures has been under debate for more than 30 years~\cite{hinch,ramaswamy}. The conceptual complexity of this collective dynamics contrasts with the formal simplicity of the (linear) Stokes equation that rules   low-Reynolds-number flows. Immersed bodies generically affect both the momentum and the mass transfers of the fluid, even when not driven by  external fields. As a result, long-range effective interactions arise between the particles due to the interplay between  the local velocity of the fluid and the motion of  the  particles.
The particle transport  problem thus corresponds to a many-body dynamical system, that  is  intrinsically non-linear. The hydrodynamic interactions actually vanish only for uniform flow fields, for which the particles would be all advected at the same speed as the fluid, irrespective of their spatial distribution. Such a condition is never achieved when the fluid is rigidly confined. For instance, when particles are driven in thin channels or in (semi)rigid films, the momentum exchange with the bounding walls causes strong distortions of the flow field, thereby inducing effective  interactions  between the particles~\cite{sackman,diamant,Wajnryb,dileonardo}. 
As it turns out, the transport of particle-ladden fluid in thin films is involved in a number of industrial and natural processes, including  protein motion in lipid membranes~\cite{sackman}, bacteria swarming~\cite{zhang,zhangr}, colloid deposition on solid surfaces~\cite{deegan1,deegan2}, and droplet-based microfluidics~\cite{seemann}. Understanding the particle transport in rigidly confined films is a first step toward the description of particle traffic in more complex  geometries such as ordered, or random porous networks.  
\begin{figure}[h!]
 \vspace*{.05in}
\begin{center}
\includegraphics[width=\columnwidth]{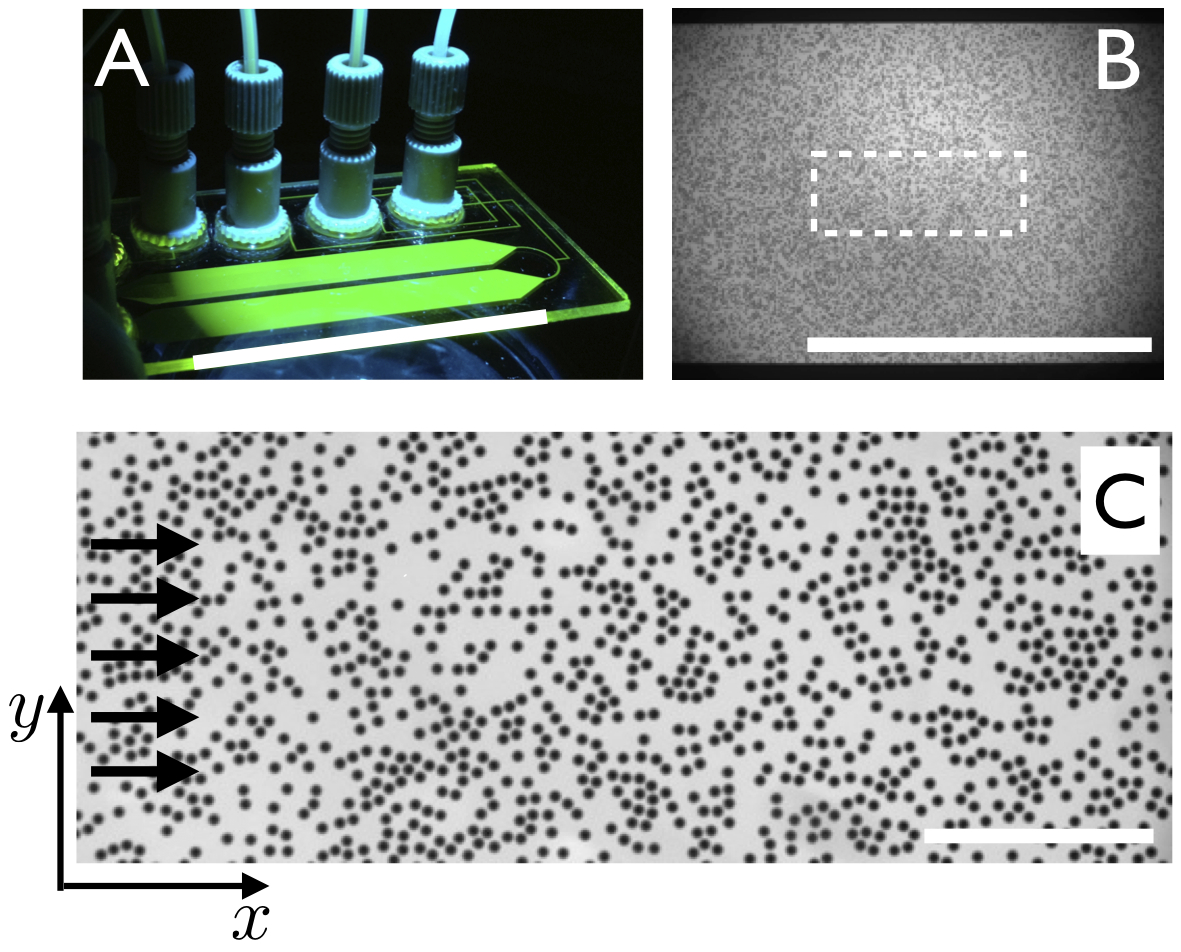}
\caption{A- Picture of the microfluidic setup. It is composed of: (i) a conventional flow-focusing junction, (ii) two dilution channels, and (iii) two identical wide observation channels. During the experiment one of these two  5-cm long channels is continuously fed with monodisperse droplets having a diameter comparable to the channel height.  Scale bar: $5$ cm. B- Close up on the main channel. Droplet diameter: $33.4\,\mu$m. Scale bar: $5$ mm. C- Typical snapshot of a movie used to track the droplet positions. The field of view corresponds to the dashed rectangle in B. The black arrows indicate the direction of the flow. Scale bar: $500\,\mu$m.}
\label{fig1}
\end{center}
\end{figure} 
The pioneering experiments by Rouyer et al.  revealed that an ensemble of confined particles advected by a uniform viscous flow displays short-range structural correlations akin to the one found in simple liquids at the molecular scale~\cite{rouyer}.  More recently,  Beatus and coworkers exploited a simple microfluidic experiment to probe the propagation of density heterogeneities
in  bidimensional emulsions~\cite{beatus}. They showed that the dynamic creation and destruction of droplet clusters also display a temporal coherence. Focusing on a semi-local quantity, the area fraction averaged over the channel width, they demonstrated that the droplet velocity is an increasing function of the local density, which is the elementary ingredient to give rise to Burgers (nonlinear) shock waves. The same phenomenology was also found in focused particle streams~\cite{champagne}. However, this elegant observation does not account for the complexity of the  spatiotemporal fluctuations generically observed at all scales in rigidly confined particle-laden fluids.

Here we combine experimental and theoretical results to shed light on the collective dynamics of  particles cruising through large-scale microfluidic channels. We first provide a comprehensive characterization of the density fluctuations observed in our experiments, and unveil their propagative nature at all scales, and in all directions. Then, 
we quantitatively demonstrate how the fluctuation spectrum of this many-body non-equilibrium system is shaped by the combination of  truly long-range hydrodynamic interactions and  local collisions.

\section{Experimental results}
Briefly, our experimental system consists of a monodisperse  emulsion flowing in a  shallow microchannel. 
 The length and width of the channel, $L\times W=5\,{\rm cm}\times 5\,{\rm mm}$, are much larger than its height, $h=27\pm0.1\,\mu \rm m$, which compares with the droplet diameter, see Fig. 1. The emulsion is therefore confined in a quasi-2D geometry.
The droplets are formed at a conventional flow-focusing junction followed by a dilution module. 
The fluid  flow-rates are imposed by   high-precision syringe pumps. Etched-glass microchips  ensure that the channel dimensions are unaffected by the flow conditions. In addition,
the geometry of the junction, and  the range of flow rates, are chosen so-that the formation of the droplet was unaffected by the dilution flow. By doing so, we accurately control both the droplet radius, $R_{\rm d}$, and the average area fraction, $\phi$, occupied by the emulsion.   We report here results obtained for $R_d=16.7\,\pm 0.3 \, \mu {\rm m}$ ($R_d/h=0.62$), and $0.21<\phi<0.56$.  Varying the droplet sizes up to $R_{\rm d}\sim 2h$ does not qualitatively change our measurements.  The droplets are visualized using fluorescence imaging.  For each experiment we tracked $\sim10^5$ particle trajectories in a region close to the center of the main channel, see Fig.~1B and~1C. More details about the microfluidic and the imaging setup are provided in the Materials and Methods section.

 In the absence of droplets, the fluid flow would be uniform along the $x$-direction in the observation region.  This is evidenced by the linear trajectories followed by  isolated droplets cruising along the channel. Conversely, even at the smallest surface fraction, when an emulsion flows, the  droplets undergo large transverse and longitudinal fluctuations in their motion, as shown in the supplementary  movie (Video S1).  These fluctuations lead to the  formation of particle clusters at all scales. Interestingly,  these clusters are clearly seen to travel at a  speed that is different from the mean droplet velocity: transverse clusters are faster, whereas the longitudinal ones are slower. However, these clusters are transient structures, they form and break apart in a continuous fashion.   Our purpose is to elucidate the physical mechanisms responsible for this complex and fluctuating dynamics. To quantify the spatiotemporal fluctuations of the  droplet density field $\rho({\bf r},t)$, where ${\bf r}=(x,y)$, we measure its power spectrum, that is the dynamic structure factor. Introducing the Fourier transform of the local density: $\rho_{\bf q,\omega'}=\frac{1}{2\pi}\int \rho({\bf r},t)e^{i({{\bf q}\cdot{\bf r}}-\omega' t)}{\rm d}{\bf r}\,{\rm dt}$, the power spectrum is defined as $|\tilde{\rho}_{{\bf q},\omega'}|^2  $, where $\tilde{\rho}({\bf r},t)\equiv\rho({\bf r},t)-\langle\rho({\bf r},t)\rangle$. 
 Practically, $\rho$ is computed from  the particle positions as $\rho({\bf r},t)\equiv \sum_i {\cal G}({\bf r}-{\bf r}_{i}(t))$, where   ${\bf r}_{i}(t)$  is  the position of the $i^{\rm th}$ droplet, and where ${\cal G}$ is a Gaussian shape function of width $R_{\rm d}/15$, see Materials and Methods.  
 \begin{figure*}
 \vspace*{.05in}
\begin{center}
\includegraphics[width=\textwidth]{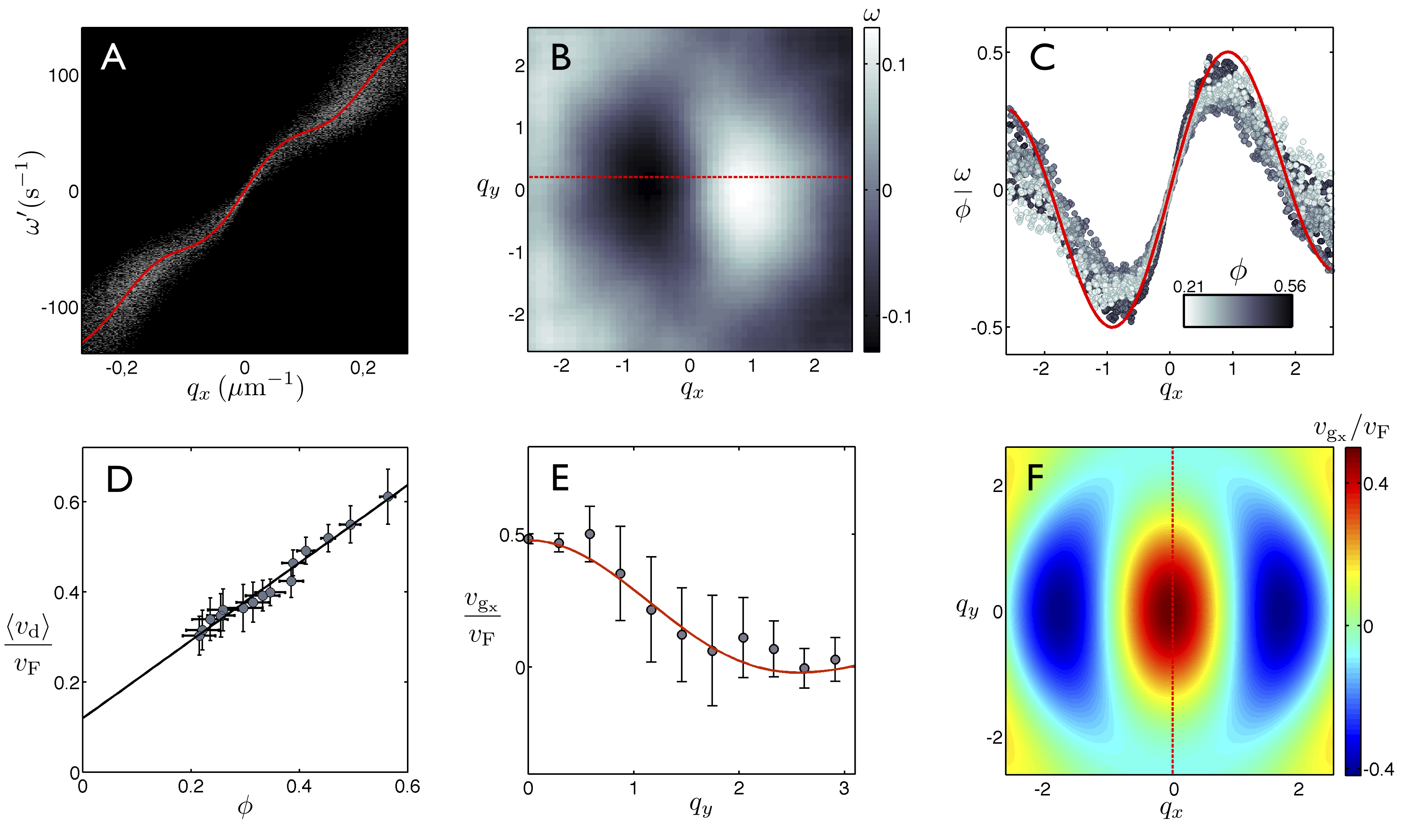}
\caption{A- Grayscale power spectrum of the density fluctuations plotted in the $(q_x,\omega')$ plane for $q_y=0.2/R_{\rm d}$. Area fraction $\phi=0.39$, mean fluid velocity: $v_{\rm F}=1\, {\rm mm}/s$. Solid line: theoretical prediction for the location of the dispersion curve. B- Experimental dispersion curve in the frame moving at $\langle {\bf v}_{\rm d}\rangle$, for $\phi=0.39$. $\omega$ is plotted versus $q_x$, and $q_y$. We recall that units are chosen so that $R_d=1$, and $v_{\rm F}=1$. The dotted line indicates the $q_y$ value corresponding to the power spectrum shown in A. C- Renormalized dispersion relations in the moving frame, for $q_y=0$. Circles: experimental data ($\phi=0.21,\,0.25,\,0.30 ,\,0.31 ,\,0.33 ,\,0.35 ,\,0.38 ,\,0.39 ,\,0.41 ,\,0.45 ,\,0.49 ,\,0.56$). Solid line: theoretical prediction, Eq.~\ref{eq8}, with no adjustable parameter. D- Variations of the (rescaled) mean droplet velocity as a function of the area fraction. Circles: experimental data. Solid line: best linear fit. The error bars account for statistical fluctuations, and correspond to the standard deviation. 
E- $v_{\rm g_x}$ plotted versus $q_y$ at $q_x=0$. Circles: experimental data for $\phi=0.56$. Solid line: Theoretical prediction with no adjustable parameter deduced from Eq.~\ref{eq8}. The error bars correspond to a 95\% confidence interval in the measurement of $v_{\rm g_x}$ from the slope of the dispersion curve.
F- Theoretical prediction for the variations of the group velocity, $v_{\rm g_x}$ with the wave vector components. Dotted line: $q_x=0$. The variations of $v_{\rm g_x}$ along this direction are shown in E. } 
\label{fig2}
\end{center}
\end{figure*}

In Fig 2A, we show a slice of a typical  power spectrum in the  $(\omega',q_x)$ plane. This example corresponds to $\phi=0.39$, and to $q_yR_{\rm d}=0.2$. Several important comments are in order:  (i) The power spectrum is localized in the Fourier space, which is the hallmark of propagative dynamics for the density fluctuations, as first noted in \cite{beatus} for the specific case of the $y$-averaged density mode ($q_y=0$). We stress that compression modes propagate even though the droplets do not interact via potential forces, and even though their inertia is negligible compared to the viscous friction at this scale. These "sound" modes originate only from the hydrodynamic coupling between the advected particles. (ii) The curve on which the spectrum is peaked corresponds to the dispersion curve of the density waves. We emphasize that it  deviates markedly from a straight line at moderate wave-lengths.  The hydrodynamic interactions do not merely renormalize the mean advection speed and cause the density fluctuations  to propagate in a dispersive fashion. (iii) The global shape of the spectrum is conserved for every area fraction, and more surprisingly for every wave vector $q_y$ provided that the wavelength remains larger than the particle size (see below).  

In all that follows, we discard the trivial non-dispersive contribution due to the advection at the mean droplet velocity $\langle {\bf v}_{\rm d}\rangle$. To do so, we focus on the density fluctuations in the frame moving at $\langle {\bf v}_{\rm d}\rangle$, and introduce the reduced pulsation  $\omega\equiv \omega'-\langle v_{\rm d}\rangle q_x$. 
Experiments done at different area fractions, and thus at different continuous phase velocities, are compared by normalizing the pulsations, and the wave vectors by $v_{\rm F}/R_{\rm d}$, and $R_{\rm d}^{-1}$ respectively. Fig 2B shows a typical dispersion relation: $\omega=\omega(q_x,q_y)$, obtained for $\phi=0.39$. The spectrum is symmetric along the $q_y$ direction as expected from the symmetry of the system. More importantly, density fluctuations propagate in all directions except in the one strictly transverse to the flow ($q_x=0$). In addition, the dispersion curve displays an axial symmetry with respect to the $q_y$-axis. It is worth noting that the sign of the associated phase velocity changes as $q_x$ increases. The long wavelength excitations propagate downstream, while the short  wavelength excitations propagate upstream. 

In Fig. 2C,  we show  that  once renormalized  by $\phi$, the dispersion relations corresponding to 12 different area fractions collapse on a single master curve. This noticeable collapse is not specific to the purely longitudinal waves and occurs for all the possible $q_y$ values. This systematic rescaling demonstrate that a unique set of physical mechanisms dictates the collective motion of the droplets,  at all scales, regardless of the droplet density.  We now establish a theoretical model which quantitatively accounts for these robust experimental findings.
\section{Theory and discussion}
We did not observe any deformation of the droplets in our experiments. Therefore,  the instantaneous configuration of the emulsion is fully determined  by the positions  of  $N$ identical axisymmetric particles: ${\bf r}_i(t)$,  $i=1\ldots N$. The dynamics of an isolated particle has proven to be correctly captured by a constant mobility coefficient, $\mu$, defined as $\dot{\bf r}_i(t)\equiv\mu {\bf v}({\bf r}_i,t)$ where ${\bf v}({\bf r}, t)$ is the in-plane fluid velocity field averaged over the channel height  in the absence of the particle $i$~\cite{beatus}. In our quasi-2D geometry, the fluid flow is a potential flow and derives from the local pressure field: ${\bf v}=-G\nabla P$ where $G=h^2/12\,\eta$, $\eta$ being the viscosity of the aqueous phase. $\bf v(\bf r,t)$ is then fully determined when considering the incompressibility condition, and the no-flux boundary conditions through the sidewalls of the channel as well. In a particle-free channel, the velocity field would be uniform, ${\bf v}=v_{\rm F}\hat{{\bf x}}$. The particles are not passive tracers ($\mu < 1$), therefore their relative motion with respect to the  fluid  results in a dipolar disturbance of the surrounding flow, as shown in~\cite{ sackman,diamant}. The potential dipolar perturbation, ${\bf v}^{\rm dip}({\bf r},{\bf r}_i(t))$, induced at the position ${\bf r}$ by a particle located at ${\bf r}_i(t)$ is defined by the modified incompressibility relation:
\begin{equation}
\nabla\cdot {\bf v}^{\rm dip}({\bf r},{\bf r}_i(t))=\sigma\partial_x\delta({\bf r}-{\bf r}_{i}(t))
\label{eq1}
\end{equation}
where $\sigma$ is the dipole strength ($\sigma>0$).
To establish the equations of motion of the $N$ particles, we now assume the dipolar disturbances to be pairwise additive. This yields:
\begin{equation}
\dot{{\bf r}}_i(t)=\mu v_{\rm F}\hat{\bf x}+\mu\sum_{j\neq i}{\bf v}^{\rm dip}({\bf r}_{i}(t),{\bf r}_j(t))
\label{eq2}
\end{equation}
We now show how to move from these $N$ coupled equations  to an hydrodynamic description for the particle density field $\rho({\bf r},t)$. We follow a well established kinetic theory framework, see e.g.~\cite{risken,menzel}. We first
introduce the $N$-point distribution function, $\rho^{(N)}({\bf r}_1, ..., {\bf r}_N,t)$, to find particles at ${\bf r}_1$,..., ${\bf r}_N
$ at the time $t$. In a stationary state $\frac{{\rm d}}{{\rm d}t}\rho^{(N)}=0$. We use this relation and  Eq.~\ref{eq1} to derive the Fokker-Planck equation obeyed by $\rho^{(N)}$:
\begin{eqnarray}
\partial_t \rho^{(N)}&+&\mu\sum_{i=1}^{N}\nabla_{{\bf r}_i}\cdot[v_{\rm F}\hat{{\bf x}}\rho^{(N)}\label{eq3}\\
&+&\sum_{j\neq i}{\bf v}^{\rm dip}({\bf r}_{i}(t),{\bf r}_j(t))\rho^{(N)} ]=0\nonumber
\end{eqnarray}

We now exploit two assumptions:  the hydrodynamic interactions are assumed to be pairwise additive, and  the particles are all identical. Therefore we can establish a mass conservation equation that couples the local density $\rho({\bf r},t)$ to the two-point distribution function $\rho^{(2)}({\bf r},{\bf r}',t)$ only. It is obtained by integrating Eq.~\ref{eq3}   over $N-1$ particle positions; 
\begin{eqnarray}
\partial_t \rho({\bf r},t)&+&\mu v_{\rm F}\partial_x \rho({\bf r},t)\label{eq4}\\
&+&\mu\nabla\cdot\int d{\bf r}'{\bf v}^{\rm dip}({\bf r},{\bf r}')\rho^{(2)}({\bf r},{\bf r}',t)=0
\nonumber
\end{eqnarray}
where,  $\rho({\bf r},t)\equiv\frac{1}{(N-1)!}\int \rho^{(N)}({\bf r},{\bf r}_2, ..., {\bf r}_N,t){\rm d}{\bf r}_2\ldots {\rm d}{\bf r}_N$, $\rho^{(2)}({\bf r},{\bf r}',t)\equiv\frac{1}{(N-2)!}\int \rho^{(N)}({\bf r},{\bf r}',{\bf r}_3, ..., {\bf r}_N,t){\rm d}{\bf r}_3\ldots {\rm d}{\bf r}_N$.
We now  assume that the particle positions decorrelate over a distance as small as one particle diameter. In addition to this mean-field approximation, we also explicitly  account for  the steric repulsion between the particles. To do so, we postulate the following closure relation for Eq.~\ref{eq4}:
\begin{equation}
\rho^{(2)}({\bf r},{\bf r}')=\left |
  \begin{array}{llr}
    0 & {\rm if} & |{\bf r}-{\bf r}'|<2R_{\rm d} \\
    \rho({\bf r})\rho({\bf r}') & {\rm if} & |{\bf r}-{\bf r}'|\ge2R_{\rm d}
  \end{array}
\right.
\label{eq5}
\end{equation} 
where $R_{\rm d}$ is the radius of a particle.   Eqs.~\ref{eq4} and \ref{eq5} define the equations of motion for
the particle-density field. In principle, the effective extent of the excluded volume could be larger that the particle radius due to short-range intermolecular repulsions, and lubrication forces. However no measurable difference with the actual droplet radius could be observed in our experiments.

We now focus on the dynamics of small density fluctuations, $\tilde{\rho}({\bf r},t)$, around an homogeneous state: $\tilde{\rho}({\bf r},t)\equiv\rho({\bf r},t)-\rho_0$, where $\rho_0=\langle \rho({\bf r},t)\rangle=\phi/(\pi R_{\rm d}^{2})$.
As done in our experiments, we  work in the frame moving at the mean droplet velocity $\langle {\bf v}_{\rm d}\rangle=\mu v_{\rm F}\hat{{\bf x}}+\mu\rho_0\int_{|{\bf r}-{\bf r}'|\ge2R_d}{\bf v}^{\rm dip}({\bf r},{\bf r}')\,{\rm d}{\bf r}'$.
At leading order in $\tilde{\rho}$, Eqs.~\ref{eq4} and \ref{eq5} then reduce to:
\begin{equation}
\partial_t\tilde{\rho}({\bf r},t)+\nabla\cdot\tilde{\bf j}({\bf r},t)=0
\label{eq6}
\end{equation}
where both the hydrodynamic interactions (long-range) and the contact interactions (short-range) are captured by the current functional $\tilde{\bf j}({\bf r},t)\equiv\mu\rho_0\int_{|{\bf r}-{\bf r}'|\ge2R_d}{\bf v}^{\rm dip}({\bf r},{\bf r}')\tilde{\rho}({\bf r}',t)\,{\rm d}{\bf r}'$.
Using Eq.~1 and focusing on particles  far from the sidewalls, we can show that $\nabla\cdot\tilde{\bf j}$ is actually a local quantity:
\begin{equation}
\nabla\cdot\tilde{\bf j}({\bf r},t)=-\frac{\mu\rho_0\sigma}{4\pi R_{\rm d}}\int \tilde{\rho}({\bf r}-2R_{\rm d}\hat{{\bf r}}')\cos\theta'{\rm d}\theta'
\label{eq7}
\end{equation}
Since $R_{\rm d}\ll W$ we have used the expression of the dipolar perturbation corresponding to an unbounded domain: ${\bf v}^{\rm dip}({\bf r},{\bf r}+2R_{\rm d}\hat{{\bf r}}')\cdot \hat{{\bf r}}'=-(\sigma\cos\theta')/8\pi R_{\rm d}^2$, where $\hat{{\bf r}}'\equiv\cos\theta'\hat{{\bf x}}+\sin\theta'\hat{{\bf y}}$~\cite{diamant}. 
We now look for plane wave solutions $\tilde{\rho}({\bf r},t)=\sum_{\bf q}\tilde{\rho}_{\bf q}\exp(i\omega t-i{\bf q} \cdot{\bf r})$ of Eq.~\ref{eq6}. After some elementary algebra we infer their dispersion relation, which is our main theoretical result:
\begin{equation}
\omega=(\mu\sigma\rho_{0})q_{x}\frac{J_{1}(2qR_{\rm d})}{2qR_{\rm d}}
\label{eq8}
\end{equation}
where $J_{1}$ is the first Bessel function. As $\omega$ is a real quantity, the above equation implies that
density waves freely propagate in the channel in qualitative agreement with our experimental observations. It is worth noting that since $\nabla\cdot\tilde{\bf j}$ is a local quantity, the form of the dispersion relation is generic, and does not depend on the channel size, and geometry. In addition, we point that $\omega$ scales as $\rho_0$, which explains the collapse of the normalized dispersion relations on a single master curve over the entire range of wave vectors, Figure~\ref{fig2}C. We now move to a quantitative comparison between our theoretical prediction and our experimental measurements. Eq.~\ref{eq8} is fully determined by two physical parameters: the droplet radius $R_d$, and  $\mu\sigma\rho_{0}$ that  quantifies the strength of the hydrodynamic couplings. To determine this latter parameter, we exploit  another specific feature of the long-range hydrodynamic interactions. In an isotropic and homogeneous system, due to their dipolar symmetry, the sum of all the hydrodynamic interactions would leave the mean droplet  velocity unchanged. However, in anisotropic channel geometries, $\langle {\bf v}_{\rm d}\rangle$ increases linearly with the mean density irrespective of the channel size~\cite{beatusrep}.   At $0^{\rm th}$ order in $\tilde \rho$,
$\langle {\bf v}_{\rm d}\rangle=\mu v_{\rm F}\hat{\bf x}+\frac{1}{2}(\mu\sigma\rho_{0})\hat{\bf x}$.
Importantly this relation provides a direct means to measure independently the last unknown parameter of our theory. Figure~\ref{fig2}D clearly shows that the measured mean droplet speed increases with $\rho_0$ in an affine manner. The strength of the hydrodynamic coupling ($\mu\sigma\rho_{0}$) is readily infered from a linear fit, Figure~\ref{fig2}D. We superimposed our theoretical predictions for the dispersion relation, Eq.~\ref{eq8}, both in the laboratory frame and in the frame moving at $\langle {\bf v}_{\rm d}\rangle$ in Figures~\ref{fig2}A and~\ref{fig2}C. We find that the agreement between the theory and the experiments is excellent over a wide range of wave vectors, and of area fractions. Without any free fitting parameters, our model quantitatively captures the dispersive nature of the density fluctuations observed in the flowing emulsions.    

To gain more physical insight into the propagation of the density waves, it is worth looking at the small-$q$ expansion of Eq.~\ref{eq8}: $\omega=\frac{1}{2}\mu\sigma\rho_{0}q_{x}[1-\frac{1}{2}(qR_{\rm d})^2]+{\cal O}\left((qR_{\rm d})^4\right)$.
At leading order, this relation is non-dispersive (linear) whatever the direction of propagation. The phase velocity scales linearly with the magnitude of the dipolar coupling $\sigma$. In addition, it does not depend explicitly on $R_d$, which implies that the small-$q$ excitations propagate only
due to the long-range hydrodynamic interactions between the particles. Conversely, the dispersive term in $\omega({\bf q})$ explicitly depends on the particle radius. At high $q$, the propagation of the density waves is set by the combination of the excluded volume interactions and the angular symmetry
of the hydrodynamic couplings. 

We close this section by recalling that one of the most striking feature observed in the flowing emulsions is the  propagation of vertical density bands which propagate significantly faster than the mean droplet flow, see supplemental material (Video S1). An homogeneous vertical band  spanning the entire width of the channel corresponds to the linear superposition of plane waves associated with  $q_y=0$, and with $q_x$s distributed around $q_x=0$.  In the frame moving at $\langle {\bf v}_{\rm d}\rangle$, their speed is  given by the $x$-component of the group velocity $v_{{\rm g}_x}(q_x,q_y)=\partial\omega/\partial{{q_x}}$ evaluated at $q=0$. In Figure~\ref{fig2}E, we plot  the experimental values of $v_{{\rm g}_x}(0,q_y)$, which we measured from the slope at the origin of the dispersion curves (as the ones shown in Figure~\ref{fig2}C). Again the agreement with the theoretical curve deduced from Eq.~\ref{eq8} is excellent. This plot reveals that the density bands extended across the entire channel width are the fastest and propagate at velocities 1.5 higher that the mean droplet flow, thereby making them highly visible on the experimental movies.
 
To go beyond this observation, we use Eq.~\ref{eq8}, to plot the magnitude of  $v_{{\rm g}_x}(q_x,q_y)$ for all $q$s, see Fig.~\ref{fig2}F. $v_{{\rm g}_x}$ displays non-monotonic variations with both $q_x$, and $q_y$ and changes its sign at high $q$s. In the frame moving at $\langle{\bf v}_{\rm d}\rangle$, and for $qR_d\ll1$, $v_{{\rm g}_x}$ is positive. The wave packets  propagate faster than the mean flow due to the hydrodynamic interactions that shape the dispersion curve in the long-wavelength limit. In contrast, the short-ranged "collisions" between the droplets result in the opposite effect: at high $q$ the density-wave packets propagate upstream ($v_{{\rm g}_x}<0$, for $qR_d>1$). 

\section{Conclusion}
In conclusion, we shed light on the spatiotemporal fluctuations generically observed in flow-driven confined suspensions.  We devised a model microfluidic experiment which made it possible to track the individual positions of hundred of thousands of identical droplets, interacting hydrodynamically, in a shallow channel.  
We demonstrated that density fluctuations freely propagate at all scales and along all directions in the channel, irrespective of  the area fraction occupied by the droplets. Introducing a  kinetic theory we elucidated the two physical ingredients that shape the dispersion curve of the linear excitations. At long wavelengths, non-dispersive sound modes propagate as a result of the dipolar hydrodynamic interactions caused by the motion of the confined droplets. Conversely, for wavelengths smaller than ten particle diameters the strongly dispersive nature of the density waves stems from the combination of both the excluded volume interactions and the  hydrodynamic  coupling. A challenging perspective to this study would be to consider the case of  self-driven particle such as rigidly confined motile cells, and chemically propelled colloids and droplets which also display intriguing coherent structures at all scales~\cite{zhang,thutupalli,ebbens}.

\begin{acknowledgments}
This work was partly funded by the Paris Emergence research program, and C'Nano Ile de France (DB). We acknowledge stimulating interactions with C. Savoie and M. Gu\'erard. We thank Bertrand Levach\'e for help with the experiments.
\end{acknowledgments}

\section{Materials and Methods}
\noindent \textit{Microfluidics and image acquisition}\\
The microfluidic device is a double etched fused silica/quartz custom-design chip  (Micronit Microfluidics). It consists of a conventional flow-focusing junction followed by a dilution module and a large channel.  We used deionized distilled water with 0.1\% SDS surfactant and fluorescein for the dispersed phase, and hexadecane (Sigma-Aldrich) for the continuous phase. The channel surface were cleaned in a UV/Ozone cleaner prior to the experiments, thereby making the glass surface highly hydrophilic. The injection into the channels is controlled by high-precision syringe pumps (Nemesys low pressure dosing modules, Cetoni). The device was mounted on a Nikon AZ100 upright macroscope. A Basler Aviator av2300-25gm (4 Megapixel, 8bit) camera was used to record the movement of the droplets in a field of view of  $2.70{\rm mm}\times 1.03{\rm mm}$ at the center of the channel. The frame rate was set at $45$Hz.
\vspace{0.5cm}\\
\textit{Data analysis}\\
From our movies, we detected the positions of the droplets to sub-pixel accuracy using a built in routine in the image processing program ImageJ. The particle velocities 
were then computed using the Matlab version of the tracking software developed by David Grier, John C. Crocker and Eric R. Weeks~\cite{crocker}. 
We used a custom Matlab routine to compute the power spectrum of the density fluctuations. First, we computed analytically the spatial Fourier transform of the 
density field at all time, using  a Gaussian shape function of width $R_{\rm d}/15$ (see main text), which typically corresponds to the precision on the particle position detection.  
Then, the temporal Fourier transform was computed numerically. To obtain the dispersion relations curves, we detected for each pulsation, $\omega$, the wave-vector corresponding
to the maximum of the power spectrum at this given pulsation.

\bibliography{basename of .bib file}

\end{document}